\documentclass[pra,aps,twocolumn,superscriptaddress,showpacs,eqsecnum,nofootinbib]{revtex4-1}
\usepackage{graphicx}
\usepackage{color}
\usepackage{hyperref}
\usepackage{amsmath}

\hypersetup{
	colorlinks = true,
	urlcolor   = blue,
	linkcolor  = red,
	citecolor  = blue
}

\begin{document}

\title{Raman sideband cooling of a single atom in an optical dipole trap:\\ Towards theoretical optimum in a three-dimensional regime}

\author{V.M. Porozova}
\affiliation{Center for Advanced Studies, Peter the Great St-Petersburg Polytechnic University, 195251, St.-Petersburg, Russia}
\affiliation{Quantum Technologies Center, M.V.~Lomonosov Moscow State University, Leninskiye Gory 1-35, 119991, Moscow, Russia}
\author{L.V. Gerasimov}
\affiliation{Faculty of Physics, M.V. Lomonosov Moscow State University, Leninskiye Gory 1-2, 119991, Moscow, Russia}
\affiliation{Quantum Technologies Center, M.V.~Lomonosov Moscow State University, Leninskiye Gory 1-35, 119991, Moscow, Russia}
\author{I.B. Bobrov}
\affiliation{Faculty of Physics, M.V. Lomonosov Moscow State University, Leninskiye Gory 1-2, 119991, Moscow, Russia}
\affiliation{Quantum Technologies Center, M.V.~Lomonosov Moscow State University, Leninskiye Gory 1-35, 119991, Moscow, Russia}
\author{S.S. Straupe}
\affiliation{Faculty of Physics, M.V. Lomonosov Moscow State University, Leninskiye Gory 1-2, 119991, Moscow, Russia}
\affiliation{Quantum Technologies Center, M.V.~Lomonosov Moscow State University, Leninskiye Gory 1-35, 119991, Moscow, Russia}
\author{S.P. Kulik}
\affiliation{Faculty of Physics, M.V. Lomonosov Moscow State University, Leninskiye Gory 1-2, 119991, Moscow, Russia}
\affiliation{Quantum Technologies Center, M.V.~Lomonosov Moscow State University, Leninskiye Gory 1-35, 119991, Moscow, Russia}
\author{D.V. Kupriyanov}
\email{kupriyanov@quantum.msu.ru}
\affiliation{Center for Advanced Studies, Peter the Great St-Petersburg Polytechnic University, 195251, St.-Petersburg, Russia}
\affiliation{Quantum Technologies Center, M.V.~Lomonosov Moscow State University, Leninskiye Gory 1-35, 119991, Moscow, Russia}

\date{\today}
\begin{abstract}
\noindent We clarify the optimal conditions for the protocol of Raman sideband cooling (RSC) of a single atom confined with a tightly focused far-off-resonant optical dipole trap (optical tweezers). The protocol ultimately pursues cooling to a three-dimensional ground state of the confining potential. We show that the RSC protocol has to fulfil a set of critical requirements for the parameters of cooling beams and the excitation geometry to be effective in a most general three-dimensional configuration and for an atom, having initial temperature between the recoil and the Doppler bounds. We perform a numerical simulation of the Raman passage for an example of an $^{85}$Rb atom taking into account the full level structure and all possible transition channels.
\end{abstract}

\pacs{42.50.Ct, 42.50.Nn, 42.50.Gy, 34.50.Rk}

\maketitle

\section{Introduction}\label{Section_I}

\noindent Progress in physics of cold atomic systems has opened an intriguing option of the optical tweezers, which allows to confine a single atom with an isolated microscopic dipole trap \cite{Grangier01,GrangierPRL02}. Recent advances of this technique allowed for near-deterministic loading of single atoms in the micro-traps \cite{GrunzweigNaturePhys10,LesterPRL15} and arrangement of tweezers into lattices of arbitrary shape with holographic techniques \cite{NogrettePRX14}. A unique advantage of the tweezers technique is the capability of dynamic transport of trapped single atoms \cite{BeugnonNaturePhys07, AhnNatureComm16} which allows one to assemble regular fully filled arrays of single atoms \cite{EndresScience16,BarredoScience16,BarredoNature18}. These atomic arrays with individual addressing and control of every site offer marvelous prospects for quantum simulation \cite{LabuhnNature16, BernienNature17} and computing \cite{SaffmanPRL15,SaffmanPRA15}. However, despite significant progress towards realizing a full set of quantum computing primitives in the neutral atom arrays \cite{SaffmanJPhysB16}, the quality of entangling two-qubit gates is still not very high even in the best state-of-the-art implementations \cite{Levine18}. One of the factors having a detrimental effect on two-qubit gate fidelity is residual motion of atoms in the trap. Temperatures of $\sim30$~$\mu$K commonly achieved in tweezers by molasses cooling result in pronounced Doppler broadening of the Rydberg excitation lasers and therefore limit the fidelity of two-qubit Rydberg gates \cite{BarredoPRA18,SaffmanJPhysB16}. Entangling gates based on local spin-exchange interactions \cite{KaufmanNature15} are even more demanding, and ideally require ground-state cooling for operation. Therefore, development of methods for cooling single neutral atoms in a microscopic dipole trap to the ground state of the confining potential are of essential importance. Fortunately, such methods for tightly confined particles do exist.

The technique of Raman sideband cooling (RSC) was originally proposed for ion traps \cite{ToschekJOSAB89,MonroePRL95,WinelandPRA90,Ghosh95} and then developed for cooling of atomic ensembles consisting of neutral atoms confined in optical lattices \cite{Zoller94,Hamann98,Vuletic98, PerrinEPL98, WeissPRL00, KermanPRL00}. In order to slow the trapped atom localized by the tweezers the RSC protocol reveals convenient and unique tool to quench its vibrational motion down to the ground state of the potential well. In several recent experimental works it was successfully applied to single neutral atoms in optical tweezers  \cite{Kaufman12,Rempe13,Thompson13,Sompet17} and for preparation of Bose-Einstein condensate \cite{Vuletic17}.

In a typical example of an alkali-metal atom the protocol would be effectively applicable once the atom is tightly confined along all three spatial directions, such that the respective oscillation frequencies are large enough to be resolved by Raman transitions between the two ground state hyperfine sublevels. Then for initially low excited vibrational modes the cooling protocol can subsequently suppress the oscillations along each major axis of the trapping potential.

In this paper we intend to analyse the capabilities of the simultaneous suppression of vibrational motion in all three oscillator's eigenmodes. A tweezer potential usually has a shape of an ellipsoidal well with two degenerate radial (transverse) and one axial (longitudinal) modes. We assume that in a typical experimental scenario the microtrap is originally loaded from an atomic ensemble, prepared in a magneto-optical trap and after a stage of molasses cooling still has a relatively high temperature and, as a consequence, a high mean vibrational number $\bar{\mathit{v}}>1$ for each mode. We are inspired and motivated by an impressive experimental progress in implementation of the RSC protocol for cooling of an alkali-metal atom in tweezers-type systems in a three-dimensional regime \cite{Kaufman12,Rempe13,Sompet17}. As we show here, in a general three-dimensional configuration there is a set of nontrivial requirements for the geometry of cooling beams and for the external parameters associated with the light pulses providing the Raman process for simultaneous suppression of all the oscillator's modes.

The paper is organized as follows: In Section \ref{Section_II} we review a general concept of the RSC protocol, describing it as a quantum process
transforming the density matrix from an initially thermal distribution down to the system ground state. Here we show that entanglement between the spin and vibrational degrees of freedom plays a crucial role for realization of the protocol in three-dimensional regime. In Section \ref{Section_III} the main element of the protocol, i.e. the stimulated Raman passage to a lower vibrational state, is clarified and optimized. This concerns geometry and requirements to the polarizations of the cooling beams  and their Rabi frequencies, and manipulation with an external magnetic field. In Section \ref{Section_IV} we illustrate the practical advantages of the suggested cooling optimization by our numerical simulations explicitly performed for an $^{85}$Rb atom. In conclusion we make some final remarks concerning the applicability of the proposed RSC design to the existing tweezers systems.

\section{Basic concept of the RSC protocol}\label{Section_II}

\noindent Consider an atom, confined in a dipole trap, as an element of a canonical thermal ensemble parameterized by temperature $T\equiv\beta^{-1}$. The atom is assumed to be optically pumped into a specific spin state (Zeeman hyperfine sublevel), which we denote as $|\mathrm{s}\rangle$ and will further address as a source state. Then the initial density matrix of the system is factorized into a product of a vibrational (thermal-equilibrium) and an internal spin (pure) parts:
\begin{eqnarray}
\hat{\rho}\!\!&=&\!\!\sum_{\mathit{v_x,v_y,v_z}}\!\!\exp\left\{\beta\left[{\cal F}(\beta)-\epsilon_{\mathit{v}_x\mathit{v}_y\mathit{v}_z}\right]\right\}%
|\mathit{v}_x,\mathit{v}_y,\mathit{v}_z\rangle\langle\mathit{v}_x,\mathit{v}_y,\mathit{v}_z|%
\nonumber\\%
&&\phantom{\sum_{\mathit{v_x,v_y,v_z}}\!\!\exp\left\{\beta\left[{\cal F}(\beta)-\epsilon_{\mathit{v}_x\mathit{v}_y\mathit{v}_z}\right]\right\}}\times|\mathrm{s}\rangle\langle\mathrm{s}|,
\label{2.1}%
\end{eqnarray}
where ${\cal F}={\cal F}(T)\equiv{\cal F}(\beta)$ is the free energy and $\epsilon_{\mathit{v}_x\mathit{v}_y\mathit{v}_z}$ is the energy of a harmonic oscillator eigenstate parameterized by the vibrational quantum numbers $\mathit{v}_x,\mathit{v}_y,\mathit{v}_z=0,1,2,\ldots$ for the $x,y,z$ major axes of the trap. We approximate the trap potential by an axially symmetric harmonic oscillator well having an axial frequency $\Omega_{\parallel}$ and a radial frequency $\Omega_{\perp}$ such that for any stationary vibrational state $|\mathit{v}_x\mathit{v}_y\mathit{v}_z\rangle$ its excitation energy is given by
\begin{equation}
\epsilon_{\mathit{v}_x\mathit{v}_y\mathit{v}_z}=\hbar\Omega_{\parallel}\left(\mathit{v}_z+\frac{1}{2}\right)+\hbar\Omega_{\perp}\left(\mathit{v}_x+\mathit{v}_y+1\right).
\label{2.2}
\end{equation}
The axial symmetry is not such a critical requirement and our discussion in its main points can be straightforwardly generalized for the case of an asymmetric trapping potential.

The central idea of the RSC protocol is to provide a sequence of stimulated Raman passages, which subsequently lower the vibrational numbers in each mode. Then in the expansion of the thermal equilibrium density matrix (\ref{2.1}) we can select the state
\begin{equation}
|\mathrm{s}\rangle\times|0,0,0\rangle\equiv|\mathrm{Dark}\rangle,%
\label{2.3}
\end{equation}
which is not affected by the Raman process and is conventionally specified as a "dark" state. The small initial population of this specific state is  enhanced step-by-step by repeating the processes, consisting of the Raman transitions and optical pumping cycles, returning the spin subsystem back to the source state. In an ideal scenario it is expected that the atom would eventually occupy the dark state with 100\% probability and therefore will be loaded into the vibrational ground state.

Let us clarify this idea by tracking the transformation of the system state at each step of the protocol. An ideal Raman passage can be expressed as a unitary transfer of a base state $|b\rangle=|s\rangle\,|\mathit{v}_x\mathit{v}_y\mathit{v}_z\rangle$, contributing into the initial density matrix (\ref{2.1}), onto a specific destination state $|d\rangle$ in accordance with the rule
\begin{eqnarray}
\lefteqn{|b\rangle\equiv|\mathrm{s}\rangle\;|\mathit{v}_x,\mathit{v}_y,\mathit{v}_z\rangle\ \ \stackrel{\mathrm{Raman}}{\Rightarrow}\ \ \ }%
\nonumber\\%
\nonumber\\%
&&\hspace{-0.5cm}C_x^{(\mathit{v}_x\mathit{v}_y\mathit{v}_z)}|\mathrm{t}_x\rangle|\mathit{v}_x\!\!-\!\!1,\mathit{v}_y,\mathit{v}_z\rangle%
+C_y^{(\mathit{v}_x\mathit{v}_y\mathit{v}_z)}|\mathrm{t}_y\rangle|\mathit{v}_x,\mathit{v}_y\!\!-\!\!1,\mathit{v}_z\rangle%
\nonumber\\%
&&+\ C_z^{(\mathit{v}_x\mathit{v}_y\mathit{v}_z)}|\mathrm{t}_z\rangle|\mathit{v}_x,\mathit{v}_y,\mathit{v}_z\!\!-\!\!1\rangle%
\nonumber\\%
\nonumber\\%
&&=\sum_{\mu=x,y,z}C_{\mu}^{(\mathit{v}_x\mathit{v}_y\mathit{v}_z)}|\mathrm{t}_{\mu}\rangle|...,\mathit{v}_{\mu}\!\!-\!\!1,...\rangle\equiv|d\rangle%
\label{2.4}%
\end{eqnarray}
We assume the Raman transfer to be a lossless dynamical process: $|C_x^{(\mathit{v}_x\mathit{v}_y\mathit{v}_z)}|^2+|C_y^{(\mathit{v}_x\mathit{v}_y\mathit{v}_z)}|^2+|C_z^{(\mathit{v}_x\mathit{v}_y\mathit{v}_z)}|^2=1$. If one of the vibrational modes, contributing into the base state, has already reached the "zero"-number then the associated amplitude factor contributing to the sum in the right-hand side should be cancelled out i.e. $C_{x}^{(0\mathit{v}_y\mathit{v}_z)}\to 0$, or $C_{y}^{(\mathit{v}_x0\mathit{v}_z)}\to 0$, or $C_{z}^{(\mathit{v}_x\mathit{v}_y0)}\to 0$. Thus the expansion (\ref{2.4}) introduces the transform in the most general configuration. The contributing target spin states $|\mathrm{t}_x\rangle$, $|\mathrm{t}_y\rangle$ and $|\mathrm{t}_z\rangle$ are unique for the chosen transition scheme and energy structure, and depend on the parameters of the Raman pulse providing such an ideal conversion.

The constructed superposition (\ref{2.4}) describes an atomic wave-packet considered at the moment of its preparation. Then at arbitrary time the state becomes time-dependent
\begin{equation}
\left.|d(t)\rangle\right|_{t>0}\!\Rightarrow\!\sum_{\mu=x,y,z}\!\!C_{\mu}^{(\mathit{v}_x\mathit{v}_y\mathit{v}_z)}|\mathrm{t}_{\mu}(t)\rangle\exp\left[-i\Omega_{\mu}t\right]|...,\mathit{v}_{\mu}\!-\!1,...\rangle,%
\label{2.5}
\end{equation}
where $\Omega_{x}=\Omega_{y}=\Omega_{\perp}$ and $\Omega_{z}=\Omega_{\parallel}$, and each spin state $|\mathrm{t}_{\mu}(t)\rangle$ has its own temporal dynamics.

The next step of the protocol consists of a non-unitary incoherent repopulation of the atom back onto the source state, which can be done with a resonant optical pumping pulse. The main requirement is that in the optical pumping process the spin density matrix should be transformed independently of the vibrational motion of the atom. That can be justified by the physical arguments based on the Lamb-Dicke effect providing that the vibrational motion is not affected by a sufficiently weak resonant pulse consisting of only few photons. Then we arrive at the separable density matrix structure similar to Eqs.~(\ref{2.1}), but with the modified part associated with the vibrational degrees of freedom. Let us think about the state (\ref{2.4}) as an entangled state expressed here by its Schmidt decomposition and consider $|\mathrm{t}_x\rangle$, $|\mathrm{t}_y\rangle$ and $|\mathrm{t}_z\rangle$ to be mutually orthogonal. If so, then after the optical pumping cycle the state of the system is reproduced in a factorized form similar to (\ref{2.1}) but with a partially enhanced population of the dark state. The details are clarified in Appendix \ref{Appendix_A}

As shown in Appendix \ref{Appendix_A}, after $n$ steps of subsequent application of Raman and optical pumping cycles the density matrix is transformed as follows:
\begin{eqnarray}
\hat{\rho}^{(n)}&=&\exp\left\{\beta{\cal F}(\beta)\right\}{\cal Z}^{(n)}(\beta)\times|0,0,0\rangle\langle0,0,0|\times|\mathrm{s}\rangle\langle\mathrm{s}|%
\nonumber\\%
&+&\ldots%
\label{2.6}%
\end{eqnarray}
where ${\cal Z}^{(n)}$ denotes the cut-off of the oscillator's partition sum containing $n$ excitations and is given by
\begin{eqnarray}
{\cal Z}^{(n)}(\beta)&=&\sum_{\begin{array}{c}\scriptstyle{\mathit{v_x,v_y,v_z}}\\ \scriptstyle{\mathit{v}_x+\mathit{v}_y+\mathit{v}_z\leq n}\end{array}}%
\exp\left[-\beta\epsilon_{\mathit{v}_x\mathit{v}_y\mathit{v}_z}\right]%
\nonumber\\%
&=&\sum_{\epsilon\leq \epsilon_n}g_{\epsilon}\exp\left[-\beta\epsilon\right].%
\label{2.7}
\end{eqnarray}
Here $g_{\epsilon}$ is the quantum degeneracy of the states $|\mathit{v_x,v_y,v_z}\rangle$ with equal energies, but different vibrational numbers, $\epsilon=\epsilon_{\mathit{v}_x\mathit{v}_y\mathit{v}_z}$, $\epsilon_n$ is the upper energy bound corresponding to $n=\max\{\mathit{v}_x+\mathit{v}_y+\mathit{v}_z\}$ excitations of the vibrational modes. Since
\begin{equation}
\lim_{n\to\infty}{\cal Z}^{(n)}(\beta)=\exp\left\{-\beta{\cal F}(\beta)\right\},%
\label{2.8}
\end{equation}
finally we arrive at
\begin{equation}
\lim_{n\to\infty}\hat{\rho}^{(n)}=|0,0,0\rangle\langle0,0,0|\times|\mathrm{s}\rangle\langle\mathrm{s}|,%
\label{2.9}
\end{equation}
such that the contribution of the other terms indicated by ellipses in (\ref{2.6}), and responsible for a residual depopulation of the dark state, vanishes with $n\to\infty$. That demonstrates the internal convergence of the process and the preparation of the system in the dark state with vanishing contribution of the excited vibrational modes.

We have constructed the basic transformations of the density matrix (\ref{2.6})-(\ref{2.9}) under an assumption that Eq.~(\ref{2.4}) uniquely reveals a Schmidt decomposition of an entangled state shared between the vibrational and the spin subsystems. But in a general case with an arbitrary Raman coupling, for which we only know that it provides a perfect transfer of the spin and the vibrational state of atom, the Schmidt decomposition of the final state could be different from (\ref{2.4}). In a most general case one can expect
\begin{eqnarray}
\lefteqn{|\mathrm{s}\rangle\times|\mathit{v}_x,\mathit{v}_y,\mathit{v}_z\rangle\ \ \stackrel{\mathrm{Raman}}{\Rightarrow}}%
\nonumber\\%
\nonumber\\%
&&=\sum_{q=1,2,3}C_{q}|\mathrm{t}_q\rangle\times|\mathrm{vib}_q\rangle,%
\label{2.10}%
\end{eqnarray}
where $C_{q}$ are the Schmidt coefficients and the dimension of the expansion coincides with the number of vibrational degrees of freedom. Overwise the protocol would access only to a part of vibrational motion and not provide the three-dimensional cooling. So the maximal entanglement is a crucial requirement for the protocol and each of the vibrational basis states $|\mathrm{vib}_q\rangle$ is given by superposition of three terms $|...,\mathit{v}_{\mu}\!-\!1,...\rangle$ with $\mu=x,y,z$. After tracing out the spin state (optical pumping cycle) the vibrational part of the density matrix transforms to a mixed state containing the coherent coupling between different modes. In accordance with (\ref{2.5}) the density matrix of the vibrational subsystem would have off-diagonal elements oscillating in time. Nevertheless, one can expect that natural dissipation of coherency, associated with any weak external perturbations of the trap potential, would soften this problem and reduce the off-diagonal components, so the basic result, given by Eqs.~(\ref{2.6})-(\ref{2.9}) is still applicable to various other (probably less optimal) Raman transition schemes. A  significant internal-state decoherence in optical tweezers has been recently observed in experiment \cite{Thompson13}.

\section{Possible scenarios of the Raman passage}\label{Section_III}

\noindent In this section we consider the possible schemes of the Raman passage suppressing the vibrational motion, which is formally expressed by Eq.~(\ref{2.4}). We show how this key element of the entire RSC protocol could be optimized depending on the thermal kinetic energy (temperature), with which the atom is initially loaded in the trap.

\subsection{The RSC for low-energy excitation of the oscillator's modes}

\begin{figure}[tp]
{$\scalebox{0.5}{\includegraphics*{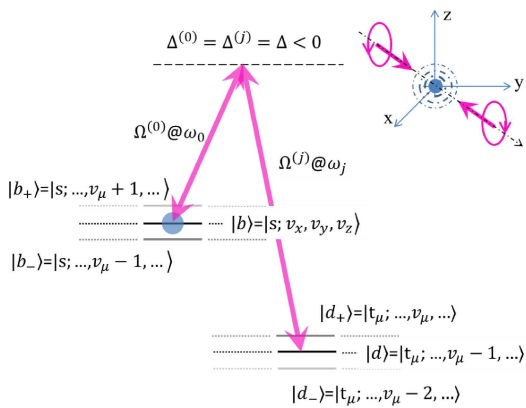}}$}
\caption{(Color online) The RSC protocol in a one-dimensional regime: the stimulated Raman scattering is initiated between the source state $|s\rangle$ and the target states $|\mathrm{t}_{\mu}\rangle$ ($\mu=x,y,z$). Each combination of the beam pairs (which direction is arbitrary shown in the diagram) could be subsequently redirected along the trap axes and with selecting the carrier frequencies $\omega_0$ and $\omega_j$ (with $j=1,2,3\Leftrightarrow\mu=x,y,z$) would provide suppression of the vibrational motion by one quantum $v_{\mu}\to v_{\mu}-1$. This constructs the Raman transfer (the main element of the RSC protocol) between the base state $|b\rangle$ and the destination state $|d\rangle$.}
\label{fig1}
\end{figure}

\noindent Let us firstly consider the situation when the dipole trap is so tight in the transverse direction that the trap oscillator is weakly excited and the mean radial vibrational numbers $\bar{v}_x\sim\bar{v}_y\sim 1$. For the axial mode we assume $\bar{v}_z>1$, but not extremely high. That physically corresponds to the condition that the atom is loaded into the tweezers with the temperature $T\sim\Omega_{\perp}$. For this particular case the RSC protocol can be separately organized for each of the vibrational degrees of freedom. A possible example of a transition scheme and an excitation geometry is shown in Fig.~\ref{fig1}, where cooling is provided by two counter-propagating collinear beams with orthogonal circular polarizations.

The combination of two beams with Rabi frequencies $\Omega^{(0)}$ and $\Omega^{(j)}$ with $j=1,2,3$ subsequently directed along the trap axes and tuned to the respective spin-vibrational transition leads to the $\Lambda$-type resonant interaction between the base state $|b\rangle$ and the destination state $|d\rangle$. For simplicity we assume the detunings from the upper hyperfine manifold in the diagram of Fig.~\ref{fig1} to be equal for all the optical fields participating in the process such that $\Delta^{(0)}=\Delta^{(j)}\equiv\Delta$. Without spontaneous losses the process can be reduced to the primitive time dynamics of the probability amplitudes $c_{b}(t)$ and $c_{d}(t)$ associated with the populations of these states:
\begin{eqnarray}
c_{b}(\tau)&=&\cos\left[\frac{1}{2}\int_0^\tau\Omega(t)dt\right]%
\nonumber\\%
c_{d}(\tau)&\sim&\sin\left[\frac{1}{2}\int_0^\tau\Omega(t)dt\right],%
\label{3.1}%
\end{eqnarray}
where $\Omega(t)$ is an effective Rabi frequency of the Raman process and in the second line we omitted an unimportant phase factor. This indicates the well-known concept of a $\pi$-pulse with $\int_0^\tau\Omega(t)dt=\pi$ as a necessary and sufficient condition to convert the atom from the level $|b\rangle$ to the level $|d\rangle$ in the two-level problem. For the cooling beams far detuned from the atomic resonance the condition can be fulfilled for sufficiently strong and long light pulses.

If after the optical pumping cycle the atom was safely repopulated back onto the $|s\rangle$-state and other disturbances, associated with the trap imperfection and cross-interaction of the vibrational modes, did not affect the entire dynamics, the atom would eventually occupy the dark state (\ref{2.3}) for a particular vibrational mode after several protocol repetitions. The process can be separately organized for each vibrational degree of freedom and will eventually cool the atom onto the dark state. But even for such a straightforward scheme it can be pointed out that at each step of the protocol the constructed $\Lambda$-type resonance would be quite sensitive to the Zeeman energy shifts in the external magnetic field, as well as to its radiation dressing (light-shifts), induced by the driving coherent fields. The importance of these effects in more complicated three-dimensional geometry will be clarified below.

\subsection{Raman passage in a three-dimensional regime}

\noindent Consider now the main scenario when the trap oscillator is excited in all three directions so that $\bar{v}_x,\bar{v}_y,\bar{v}_z > 1$ and the vibrational numbers can be relatively high. In such a situation even a rather weak cross-interaction between the vibrational modes can washout the advantage of simplicity for the mechanism of separated cooling of the trap modes described above. Some experimental protocols \cite{Kaufman12,Rempe13,Sompet17,Weiss12} have pointed out certain advantages from control over all degrees of freedom of a trapped atom.  As we show below, a more optimal strategy could be proposed to organize the simultaneous suppression of the vibrational motion in all three directions.

\subsubsection{Excitation geometry and transition scheme}

\begin{figure}[tp]
\scalebox{0.4}{\includegraphics*{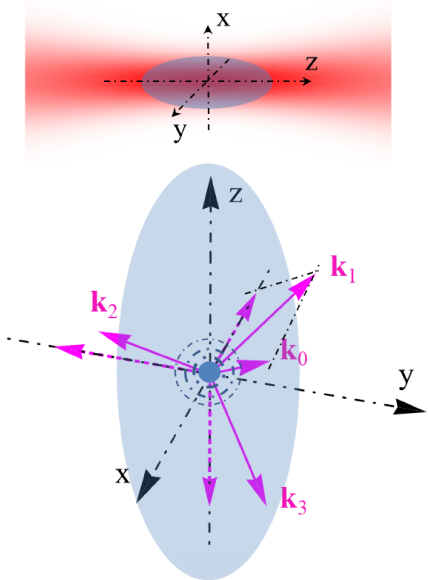}}
\caption{(Color online) Geometry of the light beams providing simultaneous quenching of the vibrational motion in the three-dimensional regime. Upper part of the figure shows the trap location in the tweezers (elliptic spot) in respect to the confining beam (red shining area) and the lower part specifies the geometry of the cooling process. The four light beams are directed along the bisectrices of the main and three adjoining octants of the trap frame. The depopulating beam with the wave vector $\mathbf{k}_0$ in combination with each of the control beams $\mathbf{k}_1$, $\mathbf{k}_2$, and $\mathbf{k}_3$ produces the recoil linear momenta  hitting the atom along the major trap axes. This is clarified by an example of a parallelogram constructed with $\mathbf{k}_0$ and $\mathbf{k}_1$.}
\label{fig2}%
\end{figure}

\noindent For three-dimensional cooling we have to generalize the scheme shown in the diagram of Fig.~\ref{fig1} up to three simultaneously acting control components with independent access for linear momentum transfer to each of the vibrational degrees of freedom. As shown in Fig.~\ref{fig2} this can be done if the wave vectors of the control beams  $\mathbf{k}_1$, $\mathbf{k}_2$, and $\mathbf{k}_3$ form isosceles triangles with the wave vector of the depopulating beam $\mathbf{k}_0$. In this excitation geometry the plane angles $\angle(\mathbf{k}_0-\mathbf{k}_1,\mathbf{k}_0)=\angle(\mathbf{k}_0-\mathbf{k}_2,\mathbf{k}_0)=\angle(\mathbf{k}_0-\mathbf{k}_3,\mathbf{k}_0)=\arccos(1/\sqrt{3})=54.7^0$ and three recoil wave vectors $\mathbf{k}_1-\mathbf{k}_0$, $\mathbf{k}_2-\mathbf{k}_0$, and $\mathbf{k}_3-\mathbf{k}_0$ become mutually orthogonal and can be directed along the major axes of the atomic trap as shown in the diagram of Fig.~\ref{fig2}, where they are visualized by dashed pink arrows. The vectors $\mathbf{k}_0,\,\mathbf{k}_1,\,\mathbf{k}_2,\,\mathbf{k}_3$ are respectively directed along the bisectrices of the main and the three adjoining octants of the trap frame. The quantization axis, coinciding with the direction of the external magnetic field, could be associated with either $\mathbf{k}_0$ (if the depopulating beam is circularly polarized) or with the beam polarization direction (if it is linearly polarized). We shall prefer the former option since the entire coupling between the base and destination states seems as more effective for this case.\footnote{In this section we discuss the RSC protocol with minimal interfere of magnetic field and light shifts with atomic energy structure. So the magnitude of the field is expected to be small and less than gauss (Earth value) and therefore can be precisely controllable in experiment. We shortly comment an alternative option in the end of Section \ref{Section_IV}.}

\begin{figure}[tp]
\scalebox{0.5}{\includegraphics*{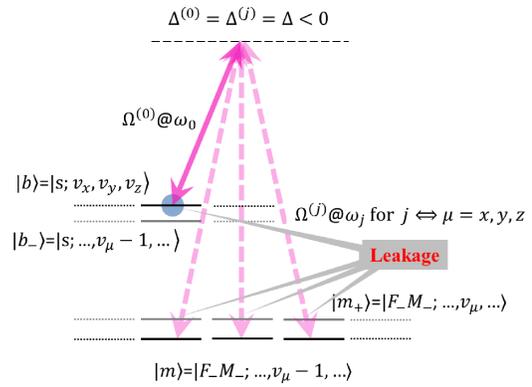}}
\caption{(Color online) The transition scheme corresponding to the excitation geometry shown in Fig.~\ref{fig2}. For a particular $j$-th control beam the Raman scattering transfers the atom onto the lower hyperfine sublevel of the ground state suppressing the respective vibrational number $\mathit{v}_{\mu}$ by one unit. Each control mode couples to a single vibrational mode such that $j=j(\mu)$. The main scattering channel competes with the off-resonant leakage onto the same spin states but with the vibrational number conserved.}
\label{fig3}%
\end{figure}

The favourable Raman process is now depicted by the transition diagram shown in Fig.~\ref{fig3}. To be specific, we will further assume that the source state $|s\rangle\equiv|F_{+},0\rangle$ is the upper state of the "clock" transition in the ground state hyperfine manifold of an alkali-metal atom. Here and throughout we will denote two possible values of the ground state spin angular momentum $F_0=F_{+},F_{-}$ with $F_{\pm}=I\pm 1/2$, where $I$ is the nuclear spin, and we will use $M_0=M_{\pm}$ for the upper/lower hyperfine sublevels. Each of the control beams opens three transition channels, such that for the $\sigma_{+}$-polarized depopulation beam the target spin states $|\mathrm{t}_x\rangle$, $|\mathrm{t}_y\rangle$, and $|\mathrm{t}_z\rangle$ are expected to be expressed as linear combinations of $|F_{-},0\rangle$, $|F_{-},1\rangle$, and $|F_{-},2\rangle$.

In accordance with the considered excitation geometry, shown in Fig.~\ref{fig2}, each $j$-th control beam with $j=1,2,3$ respectively causes the quenching of the vibrational motion along $\mu$-th axis with $\mu=x,y,z$. The partial Raman passage, associated with a particular beam, provides the transition onto the set of states $|m\rangle=|F_{-},M_{-},\ldots,\mathit{v}_{\mu}-1,\ldots\rangle$, but the complete transition accumulates the superposition of all such states with $\mu=x,y,z$. As we further explain the polarizations of the control beams should be taken as mutually orthogonal and the constructed target spin states are expected (in an ideal scenario) to slightly overlap each over, so the entire destination state $|d\rangle$ would be close to the state, optimally entangled in the spin and vibrational degrees of freedom, as we have claimed by expansion (\ref{2.4}).

Process of the stimulated resonant Raman passage from the base state $|b\rangle$ to the destination state $|d\rangle$ is attenuated by a small factor of the Lamb-Dicke parameter. If the system has unitary dynamics and the spontaneous losses are negligible, the main scattering channel would compete with the leakage associated with the weak off-resonant scattering into the same spin states but with conserved vibrational numbers. Although the leakage, also indicated in the diagram of Fig.~\ref{fig3} as a transition to the states of $|m_{+}\rangle$ group, is weak it is not accompanied by the linear momentum transfer. So the optimal RSC protocol should provide a minimal leakage per each step of the Raman passage.

\subsubsection{The time dependent Schr\"{o}dinger equation}

\noindent For a sake of notation convenience in this subsection we denote any entire electronic and vibrational state as $|m\rangle\equiv|F_0M_0;\mathit{v_x,v_y,v_z}\rangle$ for the atom in the ground state and as $|n\rangle\equiv|FM;\mathit{w_x,w_y,w_z}\rangle$ for the atom in the excited state, where we take into account that the coupling strength of the atom with the trap is different for these states. Then
the dynamical part of the Raman process is driven by the time dependent Schr\"{o}dinger equation. The wavefunction of the atom can be expanded in the basis of its stationary states as follows
\begin{equation}
|\Psi(t)\rangle=\sum_{m}c_m(t)\mathrm{e}^{-\frac{i}{\hbar}E_mt}|m\rangle+\sum_{n}c_n(t)\mathrm{e}^{-\frac{i}{\hbar}E_nt}|n\rangle,%
\label{3.2}%
\end{equation}
where the expansion coefficients (probability amplitudes) obey the Schr\"{o}dinger equation written in the energy representation:
\begin{eqnarray}
\dot{c}_n(t)&=&-\frac{i}{\hbar}\sum_m V_{nm}(t)\mathrm{e}^{i\omega_{nm}t}\,c_m(t)%
\nonumber\\%
\dot{c}_m(t)&=&-\frac{i}{\hbar}\sum_n V_{mn}(t)\mathrm{e}^{i\omega_{mn}t}\,c_n(t),%
\label{3.3}%
\end{eqnarray}
Interaction with coherent fields, considered under the standard restrictions of rotating wave approximation (RWA), is taken in the form
\begin{equation}
V_{nm}(t)= -\frac{\hbar}{2}\sum_{j=0}^{3}\Omega_{nm}^{(j)}(t)\mathrm{e}^{-i\omega_jt}%
\label{3.4}%
\end{equation}
with requirement $V_{mn}(t)=V_{nm}^{\ast}(t)$. The process is driven by four coherent fields and the matrix elements are parameterized by slowly varying time profiles of the respective overlapping pulses having different carrier frequencies. In the Schr\"{o}dinger equation this is expressed by time dependence of the Rabi frequencies defined for the depopulating component $\Omega_{nm}^{(0)}(t)$, and for the three control fields $\Omega_{nm}^{(1)}(t)$, $\Omega_{nm}^{(2)}(t)$ and $\Omega_{nm}^{(3)}(t)$ correspondingly. As we further show the system can be adjusted for one-way Raman passage of the atom from the source spin state $|\mathrm{s}\rangle$ to a set of target states $|\mathrm{t}_{\mu}\rangle$ correlated with the respective vibrational modes $|...,\mathit{v}_{\mu}\!\!-\!\!1,...\rangle$, see Eq.~(\ref{2.4}).

Eq.~(\ref{3.3}) can be simplified for far off-resonant Raman scattering so the contribution of the upper states in equations (\ref{3.2}) can be adiabatically eliminated. Indeed, by formally integrating the first line and substituting the amplitudes $c_n(t)$ into the second line we obtain
\begin{eqnarray}
\dot{c}_m(t)&=&\left(-\frac{i}{2}\right)^2\sum_{j,k}\sum_{n,m'} \Omega_{mn}^{(j)}(t)\,\mathrm{e}^{i(\omega_j-\omega_{nm})t}%
\nonumber\\%
&&\times\int_{-\infty}^{t}dt'\Omega_{nm'}^{(k)}(t')\,\mathrm{e}^{-i(\omega_k-\omega_{nm'})t'}\,c_{m'}(t').%
\nonumber\\%
\label{3.5}%
\end{eqnarray}
The integral in the right-hand side can be approximately evaluated if note that the dominating terms with $\omega_{j}-\omega_{k}+\omega_{mm'}\sim 0$ provide slow trend dynamics of the probability amplitudes on a "coarse-grained" time scale. Then the coupled modes obey the condition $\omega_j-\omega_{nm}\sim\omega_k-\omega_{nm'}$ for all possible combinations of either Rayleigh or Raman channels. In the integrand of Eq.~(\ref{3.5}) we can specify each $k$-th mode as detuned by $\Delta_{nm'}^{(k)}=\omega_k-\omega_{nm'}$ from a particular $m'\to n$ optical transition. The concept of adiabatic elimination implies that contribution of the non-exponential part of the integrand, considered as a smooth function on a time scale longer than $1/\Delta_{nm'}^{(k)}$ , can be reliably estimated by its instant value at the time given by the upper limit of the integral. Then integral in Eq.~(\ref{3.5}) can be straightforwardly evaluated and the equation takes the form
\begin{eqnarray}
\dot{c}_m(t)&=&-i\sum_{j,k}\sum_{n,m'} \frac{\Omega_{mn}^{(j)}(t)\,\Omega_{nm'}^{(k)}(t)}{4\Delta_{nm'}^{(k)}}%
\nonumber\\%
&&\times\exp\left[i(\omega_{j}-\omega_{k}+\omega_{mm'})t\right]\,c_{m'}(t),%
\label{3.6}%
\end{eqnarray}
which can be treated as a reduced time dependent Schr\"{o}dinger equation mediating the slowly varying dynamics of the probability amplitudes driven by the effective interaction Hamiltonian, defined by the right-hand side of Eq. (\ref{3.6}).

To avoid spontaneous scattering, the RSC protocol requires quite far off-resonant offset of the applied fields from the upper state hyperfine manifold, and the respective detuning should be comparable or even higher than the ground state hyperfine splitting. In this case in order to keep the process, shown in the diagram of Fig.~\ref{fig3}, as a dominated scattering channel, other competing channels of elastic Raman transitions, initiated by the control fields, should be prevented. This can be provided by the inequality $|\Omega_{mn}^{(j)}|\ll |\Omega_{nm}^{(0)}|$ for $j=1,2,3$. Then in the right-hand side of (\ref{3.6}) it is enough to leave only the terms driven by $|\Omega_{nm}^{(0)}|^2$, $\Omega_{mn}^{(j)}\,\Omega_{nm'}^{(0)}$, $\Omega_{mn}^{(0)}\,\Omega_{nm'}^{(k)}$ for $j,k=1,2,3$.

\subsubsection{The optimization scheme}

\noindent Consider the configuration with uniform rectangular profiles of the pulses, which drive the atom during the active time interval $0<t<\tau$. In accordance with the above arguments the time dependent Schr\"{o}dinger equation  can be simplified to the system of differential equations for the probability amplitudes (\ref{3.6}) by keeping only the leading resonant coupling between the upper and lower states of the Raman transition.

Now we revise our simplified notation, used in the preceding subsection, and separate again the specifications for the lower and upper hyperfine sublevels of the ground states. Let us specify those upper states, which have $M_{+}\neq 0$ and can be occupied in the entire dynamics, as $|b'\rangle\equiv|F_{+},M_{+};\mathit{v}_x\mathit{v}_y\mathit{v}_z\rangle$ with $-F_{+}\leq M_{+}\leq F_{+}$. We will distinguish the set of $|b'\rangle$-states from the base state $|b\rangle$ with $M_{+}=0$, which atom initially populates. It is expected that under optimal conditions of the Raman passage all these states are supposed to be depopulated and the atom to be converted onto the lower energy states $|m\rangle\equiv|F_{-},M_{-};\ldots,\mathit{v}_{\mu}-1,\ldots\rangle$ with $-F_{-}\leq M_{-}\leq F_{-}$. The Raman process is described by the following equations for the upper states
\begin{eqnarray}
\dot{c}_{b}(t) &=& -i\sum_{n}\frac{|\Omega_{nb}^{(0)}|^2}{4\Delta_{n}}\ \ c_{b}(t)%
\nonumber\\%
&&-i\sum_{m}\left.\left[\sum_{n}\frac{\Omega_{bn}^{(0)}\Omega_{nm}^{(j)}}{4\Delta_{n}}\right]\right|_{j=j(m)}%
\nonumber\\%
\nonumber\\%
&&\times\exp\left[i(\omega_{0}-\omega_{j}+\omega_{bm})t\right]\,c_{m}(t)+\ldots%
\nonumber\\%
\nonumber\\%
\dot{c}_{b'}(t) &=& -i\sum_{n}\frac{|\Omega_{nb'}^{(0)}|^2}{4\Delta_{n}}\ \ c_{b'}(t)%
\nonumber\\%
&&-i\sum_{m}\left.\left[\sum_{n}\frac{\Omega_{b'n}^{(0)}\Omega_{nm}^{(j)}}{4\Delta_{n}}\right]\right|_{j=j(m)}%
\nonumber\\%
\nonumber\\%
&&\times\exp\left[i(\omega_{0}-\omega_{j}+\omega_{b'm})t\right]\,c_{m}(t)+\ldots%
\label{3.7}
\end{eqnarray}
and by the complementing equations for the lower states
\begin{eqnarray}
\dot{c}_{m}(t)&=&-i\sum_{n}\frac{|\Omega_{nm}^{(0)}|^2}{4(\Delta_{n}-\Delta_{\mathrm{hpf}})}\ \ c_{m}(t)%
\nonumber\\%
&&-i\left.\left[\sum_{n}\frac{\Omega_{mn}^{(j)}\Omega_{nb}^{(0)}}{4\Delta_{n}}\right]\right|_{j=j(m)}%
\nonumber\\%
\nonumber\\%
&&\times\exp\left[i(\omega_{j}-\omega_{0}+\omega_{mb})t\right]\, c_{b}(t)%
\nonumber\\%
\nonumber\\%
&&-i\left.\sum_{b'\neq b}\left[\sum_{n}\frac{\Omega_{mn}^{(j)}\Omega_{nb'}^{(0)}}{4\Delta_{n}}\right]\right|_{j=j(m)}%
\nonumber\\%
\nonumber\\%
&&\times\exp\left[i(\omega_{j}-\omega_{0}+\omega_{mb'})t\right]\, c_{b'}(t)+\ldots%
\label{3.8}%
\end{eqnarray}
It is taken into consideration in these equations that each mode $j$ drives the atom from the state $|m\rangle$ associated with a particular trap mode, so that $j=j(m)$, as clarified in Fig.~\ref{fig3} in an example of coupling with the base state $|b\rangle$. In the right-hand side we have ordered the terms in accordance with an hierarchy of their effect on the atom's dynamics and the ellipsis indicates the neglected terms. In the denominators we have ignored the negligible difference in the field detunings so that $\Delta_{nb}=\Delta_{nm}^{(j)}=\Delta_{nm'}^{(k)}\equiv\Delta_{n}<0$ and $\Delta_{\mathrm{hpf}}>0$ denotes the hyperfine splitting in the ground state.

The first terms in the right-hand sides of Eqs.~(\ref{3.7}) and (\ref{3.8}) are associated with the light shifts of the energy states participating in the process. The subtle point is that these shifts induce undesirable dephasing to the constructed transition amplitudes, which further affects their time dynamics. The problem with light shifts can be resolved via optimal tuning of the optical modes and with proper choice of the external magnetic field. As a first step an optimal strategy suggests to tune the Raman gates in resonance with the light-shifted energy levels
\begin{equation}
\omega_{j}-\omega_{0}+\omega_{mb}+\bar{\delta}_{m}-\delta_{b}=0,%
\label{3.9}
\end{equation}
where
\begin{equation}
\delta_{b}=\sum_{n}\frac{|\Omega_{nb}^{(0)}|^2}{4\Delta_{n}}%
\label{3.10}
\end{equation}
is the light shift of the base state
and
\begin{equation}
\bar{\delta}_{m}=\sum_{n}\frac{\overline{|\Omega_{nm}^{(0)}|^2}}{4(\Delta_{n}-\Delta_{\mathrm{hpf}})}%
\label{3.11}
\end{equation}
is the mean light shift, averaged over the three final states $|F_{-},0\rangle$, $|F_{-},1\rangle$, and $|F_{-},2\rangle$, which are intended to be eventually occupied as shown in Fig.~\ref{fig3}.

Let all $b,b'$ and $m$-states be originally degenerate (zero magnetic field) and express the probability amplitudes in the form
\begin{eqnarray}
c_{b}(t) &=&\tilde{c}_{b}(t)\,\mathrm{e}^{-i\delta_{b}t}%
\nonumber\\%
c_{b'}(t) &=&\tilde{c}_{b'}(t)\,\mathrm{e}^{-i\delta_{b}t}%
\nonumber\\%
c_{m}(t) &=&\tilde{c}_{m}(t)\,\mathrm{e}^{-i\bar{\delta}_{m}t}%
\label{3.12}
\end{eqnarray}
and substitute them in Eqs.~(\ref{3.7}) and (\ref{3.8}). Then Eqs.~(\ref{3.7}) transforms to
\begin{eqnarray}
\dot{\tilde{c}}_{b}(t) &=& -i\sum_{m}\left.\left[\sum_{n}\frac{\Omega_{bn}^{(0)}\Omega_{nm}^{(j)}}{4\Delta_{n}}\right]\right|_{j=j(m)}\,\tilde{c}_{m}(t)+\ldots%
\nonumber\\%
\nonumber\\%
\dot{\tilde{{c}}}_{b'}(t) &=& -i\left[\sum_{n}\frac{|\Omega_{nb'}^{(0)}|^2}{4\Delta_{n}}-\delta_b\right]\ \tilde{c}_{b'}(t)%
\nonumber\\%
&&-i\sum_{m}\left.\left[\sum_{n}\frac{\Omega_{b'n}^{(0)}\Omega_{nm}^{(j)}}{4\Delta_{n}}\right]\right|_{j=j(m)}\,\tilde{c}_{m}(t)+\ldots%
\nonumber\\%
\label{3.13}
\end{eqnarray}
and Eq.~(\ref{3.8}) transforms to
\begin{eqnarray}
\dot{\tilde{c}}_{m}(t)&=&-i\left[\sum_{n}\frac{|\Omega_{nm}^{(0)}|^2}{4(\Delta_{n}-\Delta_{\mathrm{hpf}})}-\bar{\delta}_m\right] \ \tilde{c}_{m}(t)%
\nonumber\\%
&&-i\left.\left[\sum_{n}\frac{\Omega_{mn}^{(j)}\Omega_{nb}^{(0)}}{4\Delta_{n}}\right]\right|_{j=j(m)}\, \tilde{c}_{b}(t)%
\nonumber\\%
\nonumber\\%
&&-i\left.\sum_{b'\neq b}\left[\sum_{n}\frac{\Omega_{mn}^{(j)}\Omega_{nb'}^{(0)}}{4\Delta_{n}}\right]\right|_{j=j(m)}\, \tilde{c}_{b'}(t)+\ldots%
\nonumber\\%
\label{3.14}%
\end{eqnarray}
As verified by our numerical simulations, condition (\ref{3.9}) already provides the effective Raman repopulation but not the required entanglement between vibrational and spin subsystems. That is because the Raman transition is most effective to $|F_{-},1\rangle$ state, which has the best resonance coupling when the light-induced Zeeman splitting is comparable with the average light shift. The importance of entanglement was clarified in our remark in the end of Section \ref{Section_II}. The situation can be resolved if the residual $m$-dependent light shifts of the target spin states, contributing to the first line of Eq.~(\ref{3.14}), would be  eliminated by applying an external magnetic field forming a compensating linear slope of the Zeeman energy levels. Further we will assume such a specially prepared degenerate structure of states $|F_{-},0\rangle$, $|F_{-},1\rangle$, and $|F_{-},2\rangle$.

Now we turn to the discussion of the mutual beam geometry, shown in the diagram of Fig.~\ref{fig2}, and are concerned how to arrange the optimal polarization choice of the control modes. If we look at the basic strategy of the RSC-protocol it would seem optimal that the process was approached by an effective two-level transition between the base state $|b\rangle$ and the destination state $|d\rangle$, defined as a superposition of the target spin and vibrational states, as explained by Eq.~(\ref{2.4}). If we make a few iteration steps with equations (\ref{3.13}) and (\ref{3.14}) we can construct the perturbation theory expansion for the system evolution operator. Each contribution to this expansion consists of the product of the effective Hamiltonian matrix elements, listed in Appendix \ref{Appendix_B}, and we obtain that the isolated interplay between states $|b\rangle$ and $|d\rangle$ to be realized if (i)
\begin{equation}
\Omega^{(1)}\eta_{\perp}\sqrt{\mathit{v}_x}=\Omega^{(2)}\eta_{\perp}\sqrt{\mathit{v}_y}=\Omega^{(3)}\eta_{\parallel}\sqrt{\mathit{v}_z}
\label{3.15}
\end{equation}
where $\Omega^{(j)}$ denotes the reduced Rabi frequency for the $j$-th mode, see (\ref{b.14}); and if (ii) all the polarizations of the control modes are mutually orthogonal. If both these conditions are fulfilled, the atom's dynamics becomes unitary (and periodic in time) swapping where the state $|b\rangle$ becomes destination state for the base state $|d\rangle$ and vice versa, similarly to (\ref{3.1}).

\begin{figure}[tp]
\scalebox{0.5}{\includegraphics*{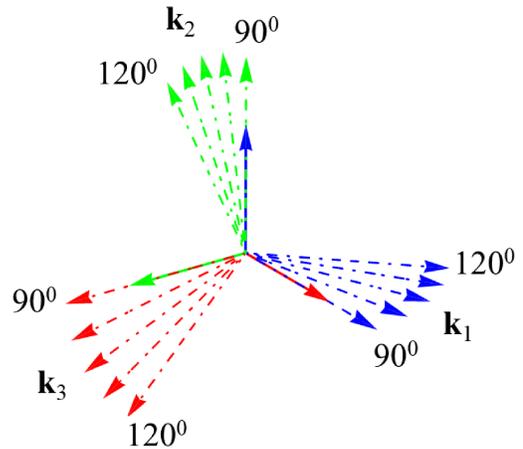}}
\caption{(Color online) A diagram showing how three light beams with mutually orthogonal polarizations, indicated by solid arrows of different colors, would have symmetric, but generally non-orthogonal, propagation directions along the wave vectors $\mathbf{k}_1$, $\mathbf{k}_2$, $\mathbf{k}_3$ with mutual angles varied from $90^0$ to $120^0$, see text for detailed comments.}
\label{fig4}%
\end{figure}

The orthogonality of polarizations can be justified for the proposed excitation geometry as proven by the diagram shown in Fig.~\ref{fig4}. In this figure the directions of the three control beams (non-orthogonal themselves!), shown by dashed arrows, are rotated around the three fixed orthogonal polarization directions, shown by solid arrows. The required mutual angle between the beams of each pair is $109.5^0$ (see Fig.~\ref{fig2}) and is just inside the accessible interval varying from $90^0$ to $120^0$.

The system of differential equations (\ref{3.13}) and (\ref{3.14}) can be numerically solved and its solution gives us the target spin states contributing to the decomposition (\ref{2.4})
\begin{eqnarray}
C_x^{(\mathit{v}_x\mathit{v}_y\mathit{v}_z)}|\mathrm{t}_x\rangle\!&=&\!\sum_{M_{-}}\left.c_{m}(\tau)\right|_{m=F_{-},M_{-};\mathit{v_x}-1,\mathit{v_y,v_z}}\left|F_{-},M_{-}\right\rangle%
\nonumber\\%
C_y^{(\mathit{v}_x\mathit{v}_y\mathit{v}_z)}|\mathrm{t}_y\rangle\!&=&\!\sum_{M_{-}}\left.c_{m}(\tau)\right|_{m=F_{-},M_{-};\mathit{v_x},\mathit{v_y}-1,\mathit{v_z}}\left|F_{-},M_{-}\right\rangle%
\nonumber\\%
C_z^{(\mathit{v}_x\mathit{v}_y\mathit{v}_z)}|\mathrm{t}_z\rangle\!&=&\!\sum_{M_{-}}\left.c_{m}(\tau)\right|_{m=F_{-},M_{-};\mathit{v_x},\mathit{v_y},\mathit{v_z}-1}\left|F_{-},M_{-}\right\rangle.%
\nonumber\\%
\label{3.16}
\end{eqnarray}
If the source state is taken as the upper state of the "clock"-transition  $|s\rangle=|F_{+},0\rangle$ then the target spin states $|\mathrm{t}_x\rangle$, $|\mathrm{t}_y\rangle$, and $|\mathrm{t}_z\rangle$ would be preferably superposed as linear combinations of $|F_{-},0\rangle$, $|F_{-},1\rangle$, and $|F_{-},2\rangle$ in accordance with the favorable process shown in Fig.~\ref{fig3}.

Such an ideal scenario is actually disturbed by an imperfection associated with violation of the condition (\ref{3.15}), which cannot be fulfilled for all the vibrational states, and as a consequence with partial occupation of $|b'\rangle$ states. Furthermore the processes of off-resonant Raman conversion onto the states $|m_{+}\rangle$ preserving the atom's vibrational mode creates leakage from the main scattering channel, see Fig.~\ref{fig3}. The respective amplitude can be estimated with the aid of the perturbation theory technique
\begin{equation}
c_{m_{+}}(\tau)\sim -i\int_0^{\tau}dt\sum_{j=1}^{3}\left[\sum_{n}\frac{\Omega_{m_{+}n}^{(j)}\Omega_{nb}^{(0)}}{4\Delta_{n}}\right]\mathrm{e}^{i\Omega_{\mu}t}\, c_{b}(t),%
\label{3.17}%
\end{equation}
where $c_{b}(t)$ is approximated by the solution of Eq.~(\ref{3.13}) substituted into Eq.~(\ref{3.12}). Here $\Omega_{\mu}$, with $\mu=\mu(j)$, is the vibrational frequency of the trap oscillator: $\Omega_x=\Omega_y=\Omega_{\perp}$ and $\Omega_z=\Omega_{\parallel}$. If the pulse duration is long enough such that $\Omega_{\mu}\tau\gg 1$, the oscillating factor in the integrand suppresses the integral and reduces the amplitude $c_{m_{+}}(\tau)$.

The minimal pulse duration, for which the competing off-resonant process could be safely ignored, requires that the off-resonance reduction factor  $1/\Omega_{\mu}\tau\ll 1$ competes with and should be smaller than the Lamb-Dicke factor $\eta_{\mu} = k_0\,\sqrt{\hbar/2m\Omega_{\mu}}\ll 1$, suppressing the transition amplitude of the Raman passage. The sufficient condition $1/\Omega_{\mu}\tau\ll \eta_{\mu}$ indicates that the pulse duration $\tau$ should be taken longer than $k_0^{-1}/\sqrt{\hbar\Omega_{\mu}/2m}\sim \lambdabar_0/\delta\mathrm{v}_\mu$, where $\lambdabar_0=k_0^{-1}$ and $\delta\mathrm{v}_\mu=\sqrt{\hbar\Omega_{\mu}/2m}$ is the quantum uncertainty of the atomic velocity in the ground state of the oscillator well in $\mu=x,y,z$-directions. This estimate shows us that the trap potential confining the atom should be sufficiently tight. The structure of the matrix elements presented in Appendix \ref{Appendix_B} softens the above requirement by a factor of $\sqrt{\mathit{v}_\mu}$, which in the estimate changes $\delta\mathrm{v}_\mu$ by a thermal velocity in the trap so for typical initial temperature, associated with preliminary cooling in the magneto-optical trap, the pulse duration should be about \textit{microseconds} or longer. Furthermore we obtain another inconvenient problem: for the light pulses, having such a long duration, the Raman-passage should be prevented against the losses associated with the spontaneous incoherent scattering, which were ignored in our analysis.

\section{Numerical simulations}\label{Section_IV}
\noindent To illustrate the above arguments we present a set of numerical simulations, which were done for an example of ${}^{85}$Rb with the source state $|F_{+},0\rangle=|3,0\rangle$ as shown in Fig.~\ref{fig3}. We assume the mode frequencies $\Omega_{\perp}=2\pi\times 200$~kHz and $\Omega_{\parallel}=2\pi\times 100$~kHz for the trap oscillator, which are potentially attainable parameters for optical tweezers \footnote{Here the confinement strengths in the radial and axial directions are taken in proportion $2:1$ to provide a sufficiently small value of the Lamb-Dicke parameter as important constraint to the RSC protocol. However the ratio within $(4\div 7):1$ between radial and axial mode frequencies are typically observed in experiments. Surprisingly but in experiments, see \cite{Kaufman12} as example, the heating during the optical pumping cycle is not so critical and axial cooling is attainable for a quite shallow trap with $\Omega_{\parallel}\sim 2\pi\times 30$~kHz.}. That makes the Lamb-Dicke parameters $\eta_{\perp}\sim 0.13$ and $\eta_{\parallel}\sim 0.18$ respectively for the radial and axial modes. With a relatively low initial temperature about 20~$\mu$K the mean values of the vibrational numbers for a trapped rubidium atom can be estimated as $\bar{v}_{\perp}\sim 2$ and $\bar{v}_{\parallel}\sim 4$. The Raman passage should be optimized by the condition (\ref{3.15}) based on the mean vibrational numbers and we present our calculations for $v_{\perp}=\bar{v}_{\perp}$ and $v_{\parallel}=\bar{v}_{\parallel}$ as well as for the numbers shifted from their mean values by one standard deviation. To uncover the evidence of radiation dressing of the atomic states and its interference with the transition dynamics, we have made comparative simulations of the Raman passage for two optical detunings from the excited state hyperfine manifold of $\Delta=-1000\gamma$ and $\Delta=-5000\gamma$. For the reduced Rabi frequencies, defined by (\ref{b.14}), we chose $\Omega^{(0)}=20\,\gamma$, $\Omega^{(1)}=\Omega^{(2)}=\gamma$, and $\Omega^{(3)}$ is fixed by condition (\ref{3.15}).

Although the used parameters would let us consider the reduced version of the Schr\"{o}dinger equation (\ref{3.13}) and (\ref{3.14}) with a cut-off, we performed our simulations based on the Schr\"{o}dinger equation in its general form (\ref{3.6}), keeping all the contributions. That lets us track the system dynamics toward its long term asymptote. We have assumed that an additional external magnetic field, compensating the slope of the light shifts, had provided the Zeeman degeneracy for the lower hyperfine sublevel. In turn that makes a partial compensation of the slope among the Zeeman states of the upper hyperfine sublevel as well. But the latter is only approximately achieved because of the difference in the frequency denominators in Eqs.~(\ref{3.10}) and (\ref{3.11}).

\begin{figure}[tp]
\scalebox{0.5}{\includegraphics*{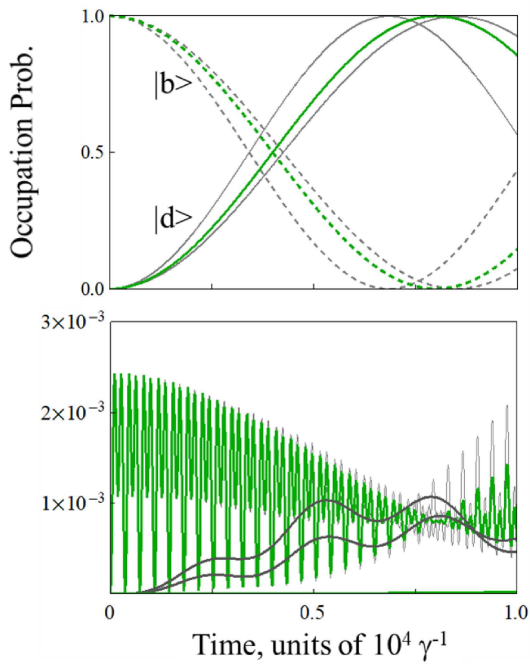}}
\caption{(Color online) \emph{Upper plot}: time dependence of the occupation probabilities for the destination states $\sum_{m}|c_m(t)|^2$ (solid curves), and for the base states $|c_b(t)|^2$ (dashed curves) calculated for the detuning from optical resonance $\Delta=-1000\gamma$ and for varied vibrational quanta. The green curves indicate the Raman passage for $v_{\perp}=\bar{v}_{\perp}$ and $v_{\parallel}=\bar{v}_{\parallel}$ and the satellite curves (gray) show the bounds of its possible variation within one standard deviation of the vibrational numbers from they mean values, see text for details. Other calculation parameters are specified in the text. All the functions demonstrate periodic dependence and the system evolves to the destination state at the extremal points. The time point, referring to complete depopulation of the base state, determines the pulse duration $\tau$ for the optimal Raman passage. \emph{Lower plot}: the imperfection of the process $\sum_{b'\neq b}|c_{b'}(t)|^2$ and leakage from the main channel $\sum_{m_{+}}|c_{m_{+}}(t)|^2$, shown for the same set of calculation parameters. Occupations of the $|b'\rangle$ states are expressed by smooth dependencies and for the optimal parameters are negligible, so the respective dependence is unresolved in the graph scale. Weak leakage to $|m_{+}\rangle$ states demonstrates an oscillatory behavior with an amplitude less than one percent.}
\label{fig5}%
\end{figure}

In Fig.~\ref{fig5} we reproduce the time dependence of $|c_b(t)|^2$ (occupation of the base state),  $\sum_{m}|c_m(t)|^2$ (occupation of the destination state), and $\sum_{b'\neq b}|c_{b'}(t)|^2$ (the repopulation imperfection) together with $\sum_{m_{+}}|c_{m_{+}}(t)|^2$ (leakage from the main scattering channel). These graphs, plotted for detuning $\Delta=-1000\gamma$, confirm that the system indeed has a tendency to complete depopulation of the base state $|b\rangle$ towards its conversion onto the destination state $|d\rangle$, created at the extremal points of the slightly distorted periodical dynamics. The efficiency of the process is quite persistent to reasonable variations of the vibrational numbers that justifies the conversion of the entire cooling protocol down to the dark state. In the figure we show the maximal amplitude of such variations, which we obtained with either adding or subtracting one standard deviation to each mean vibrational number.

It should be pointed out here that for detuning $\Delta=-1000\gamma$ despite of the compensating magnetic field the Zeeman structure of the upper hyperfine sublevel is still resolved, so that only the main resonance optical transitions, shown in Fig.~\ref{fig3}, are actually involved in the Raman process. The optimal condition (\ref{3.15}) is not critical for this case and the effective transfer is possible for any initial vibrational state. However, as can be seen from the dependencies plotted in Fig.~\ref{fig5}, the occupation of the $|d\rangle$-state is attained at different moments of time for different initial conditions.

The leakage demonstrates a periodical dependence as well, but with a strong oscillatory behavior inside the main period. That results from a weak coherent coupling of the base and $|m_{+}\rangle$ states, expressed by a small and fast oscillating transition probability, similarly to the Rabi-type oscillations in a two-level problem. The occupation of these states tends to vanish at those moments of time when the $|b\rangle$ state becomes depopulated. The imperfection gives weaker contribution than leakage and both the processes are practically negligible within the made approximations and within the validity range of our model.

\begin{figure}[tp]
\scalebox{0.5}{\includegraphics*{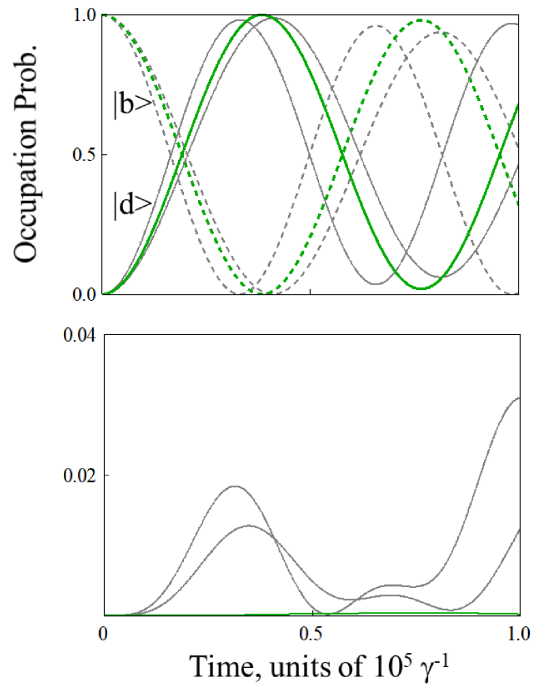}}
\caption{(Color online) Same as in Fig.~\ref{fig5} but for the detuning $\Delta=-5000\gamma$. In the optimal regime the imperfection makes negligible contribution to the population dynamics, but the $|b'\rangle$ states become occupied within a few percent of probability for deviations of the vibrational numbers from their mean values. The leakage (lower panel) is invisible in the graph scale.}
\label{fig6}%
\end{figure}

In Fig.~\ref{fig6} we show the time dynamics of the occupation probabilities for the Raman passage with detuning $\Delta=-5000\gamma$. The main features of the dynamics are the same as in the previous example, but the process becomes more sensitive to the optimal conditions (\ref{3.15}). That is a direct consequence of vanishing Zeeman splitting in the upper hyperfine sublevel. So it could be expected that in the case of $-\Delta\gg\Delta_{\mathrm{hpf}}\sim 500\,\gamma$ the imperfection would affect on efficiency of the RSC protocol. Nevertheless, as can be seen from the plotted dependencies, for the considered calculation parameters the imperfection is still small and ignorable in the general population balance.

As we emphasized in Section \ref{Section_II}, in order to optimize the RSC protocol in a three dimensional regime it would be desirable that the target spin states $|\mathrm{t}_x\rangle$, $|\mathrm{t}_y\rangle$, and $|\mathrm{t}_z\rangle$ would be mutually orthogonal. This requirement is not critical but at least these states should be prepared as linearly independent in the spin subspace. We have verified that in the considered example the target states overlaps are $|\langle \mathrm{t}_x|\mathrm{t}_y\rangle|\sim|\langle \mathrm{t}_x|\mathrm{t}_z\rangle|\sim|\langle \mathrm{t}_y|\mathrm{t}_z\rangle|\sim 0.66$ such that the constructed states provide a complete but non-orthogonal basis in the spin subspace. Then the destination state $|d\rangle$ indeed forms a maximal entangled state as it is required for entire conversion of the cooling process.

Here we can point out that with anticipating a more complicated experimental design the orthogonality of the target spin states can be provided by preparation of the control beams having different carrier frequencies. Imagine that the Zeeman sublevels, shown in Fig.~\ref{fig3}, are split by an external magnetic field such that they are perfectly resolved for the driving lasers. In this case if each of the control beams was tuned in resonance for a particular  spin transition specifically selected for each vibrational mode then the atom could be repopulated to different specific Zeeman states for different vibrational modes. The problem in realization of such scenario is in additional technical difficulties of the precise magnetic field control and in preparation of three strobe-type laser pulses with different carrier frequencies precisely resolved in MHz spectral domain.

\section{Conclusion}

\noindent In the paper we have presented a theoretical analysis of the RSC protocol, which is convenient and commonly used experimental tool for control and quenching of the vibrational motion of a single neutral atom confined with a dipole trap (optical tweezers). As we have shown and highlighted throughout our discussion the simultaneous control of all the degrees of freedom requires that in optimal configuration the spin states and vibrational modes of the atom would be entangled after the Raman cycle of the protocol. Next observation is in a nontrivial geometry associated with the excitation scheme adjusted for individual control of each major vibrational mode. To make the Raman passage most effective the parameters of the control pulses obey the set of critical requirements, which concern the control beam polarizations and reduced Rabi frequencies, see (\ref{3.15}). As we have shown, with relevant choice of the external magnetic field and cooling beams the Raman process can be transformed to an effective two-level transition scheme. That is supported by our theoretical estimates, presented in Section \ref{Section_III}, and by numerical simulations, presented in Section \ref{Section_IV}, which are connected with optimization of the external parameters for RSC experiments with alkali-metal atoms.

As follows from our numerical results shown in Figs.~\ref{fig5},~\ref{fig6} and associated with the original equilibrium thermal distribution the duration of the Raman pulse has a slight dependence on the initial number of quanta in the vibrational state, which were varied near the mean equilibrium values. As the protocol proceeds the vibrational state evolves and the average number of quanta reduces, the statistical distribution of the state evolves away from the equilibrium Gibbs measure. Further optimization of the cooling protocol may require changing the pulse length of the Raman pulses to sustain optimal Raman passage conditions for later stages of the cooling procedure. This possibility will be explored elsewhere.

Let us note, that the optimal geometry of Raman beams, shown in Fig.~\ref{fig2} and associated with the paraxial approximation, is not so straightforward to implement in a real experimental setup. Tight focusing is required to obtain sub-micron sized dipole traps, which are a prerequisite for single atom trapping in the collisional blockade regime. There are certain diificulties for experimental preparation of a dipole trap, which would be sufficiently tight in axial direction. One usually needs high-numerical-aperture lenses or objectives to achieve sub-micron waists for the tweezers. Typically numerical-aperture more than $0.5$  is used, which means that a significant part of the full solid angle will be covered by the lens itself, necessarily restricting the possible angles between the Raman beams and the trap axes. Nevertheless, we believe, that it is important to consider the geometry described here, since it corresponds to a theoretical optimum.

\section*{Acknowledgements}
\noindent This work was supported by the Russian Foundation for Basic Research under Grants \# 18-02-00265-A and \#19-52-15001-CNRS-a, by the Russian Scientific Foundation under Grant \# 18-72-10039, and by the Foundation for the Advancement of Theoretical Physics and Mathematics "BASIS" under Grant \# 18-1-1-48-1.

\appendix
\section{Transformation of the density matrix}\label{Appendix_A}

\noindent In this appendix we clarify the mathematical details of the density matrix transformation and verify the sufficient conditions, which provide the key result (\ref{2.6})-(\ref{2.9}) introduced in the main text.

After the atom makes a Raman passage (\ref{2.4}) its total density matrix (\ref{2.1}) is modified as follows
\begin{eqnarray}
\lefteqn{\hat{\rho}\Rightarrow\exp\left\{\beta\left[{\cal F}(\beta)-\epsilon_{000}\right]\right\}|0,0,0\rangle\langle0,0,0|\!\times\!|\mathrm{s}\rangle\langle\mathrm{s}|}%
\nonumber\\%
\nonumber\\%
&&\hspace{-0.5cm}+\!\!\!\!\!\!\sum_{\begin{array}{c}\scriptstyle{\mathit{v_x,v_y,v_z}}\\ \scriptstyle{\mathit{v}_x+\mathit{v}_y+\mathit{v}_z\geq 1}\end{array}}%
\!\!\!\!\!\!\exp\left\{\beta\left[{\cal F}(\beta)-\epsilon_{\mathit{v}_x\mathit{v}_y\mathit{v}_z}\right]\right\}%
\sum_{\mu=x,y,z}\sum_{\mu'=x,y,z}%
\nonumber\\%
\nonumber\\%
&&C_{\mu}^{(\mathit{v}_x\mathit{v}_y\mathit{v}_z)}C_{\mu'}^{(\mathit{v}_x\mathit{v}_y\mathit{v}_z)\ast}%
|\mathrm{t}_{\mu}\rangle|...,\mathit{v}_{\mu}\!\!-\!\!1,...\rangle\langle...,\mathit{v}_{\mu'}\!\!-\!\!1,...|\langle \mathrm{t}_{\mu'}|%
\nonumber\\%
\label{a.1}%
\end{eqnarray}
and expresses a non-separable mixed state further evolving in accordance with Eq.~(\ref{2.5}).

An incoherent optical pumping cycle breaks the time dynamics and converts the system again to the steady state separable product of a mixed vibrational component and a pure spin component
\begin{equation}
\rho^{(1)}=\rho_{\mathrm{vib}}\times|\mathrm{s}\rangle\langle\mathrm{s}|
\label{a.2}
\end{equation}
For a tight trap with a sufficiently small Lamb-Dicke parameter the optical pumping process does not affect the vibrational degrees of freedom and has the only effect of repopulating the atom onto the source spin state $|s\rangle$, such that the density matrix $\rho_{\mathrm{vib}}$ is the same before and after the repopulation. If the target states $|\mathrm{t}_x\rangle$, $|\mathrm{t}_y\rangle$ and $|\mathrm{t}_z\rangle$ are mutually orthogonal, we obtain
\begin{eqnarray}
\hat{\rho}_{\mathrm{vib}}&=&\mathrm{Tr}_{\mathrm{spin}}'\hat{\rho}%
\nonumber\\%
&=&\exp\left\{\beta\left[{\cal F}(\beta)-\epsilon_{000}\right]\right\}|0,0,0\rangle\langle0,0,0|
\nonumber\\%
\nonumber\\%
&&\hspace{-0.5cm}+\!\!\!\!\!\!\sum_{\begin{array}{c}\scriptstyle{\mathit{v_x,v_y,v_z}}\\ \scriptstyle{\mathit{v}_x+\mathit{v}_y+\mathit{v}_z\geq 1}\end{array}}%
\!\!\!\!\!\!\exp\left\{\beta\left[{\cal F}(\beta)-\epsilon_{\mathit{v}_x\mathit{v}_y\mathit{v}_z}\right]\right\}%
\sum_{\mu=x,y,z}|C_{\mu}^{(\mathit{v}_x\mathit{v}_y\mathit{v}_z)}|^2
\nonumber\\%
\nonumber\\%
&&\times |...,\mathit{v}_{\mu}\!\!-\!\!1,...\rangle\langle...,\mathit{v}_{\mu}\!\!-\!\!1,...|,%
\label{a.3}%
\end{eqnarray}
where in the second term of the right-hand side we can select the ground vibrational state (i. e. the contributions with $\mathit{v}_x+\mathit{v}_y+\mathit{v}_z=1$) and add it to the first term. So we get
\begin{eqnarray}
\lefteqn{\hat{\rho}^{\mathrm{vib}}=\exp\left\{\beta\left[{\cal F}(\beta)-\epsilon_{000}\right]\right\}}%
\nonumber\\%
&&\times\left\{1+2\exp[-\beta\hbar\Omega_{\perp}]+\exp[-\beta\hbar\Omega_{\parallel}]\right\}%
|0,0,0\rangle\langle0,0,0|
\nonumber\\%
\nonumber\\%
&&+\!\!\!\!\!\!\sum_{\begin{array}{c}\scriptstyle{\mathit{v_x,v_y,v_z}}\\ \scriptstyle{\mathit{v}_x+\mathit{v}_y+\mathit{v}_z\geq 1}\end{array}}%
\!\!\!\!\!\!\exp\left\{\beta\left[{\cal F}(\beta)-\epsilon_{\mathit{v}_x\mathit{v}_y\mathit{v}_z}\right]\right\}%
\nonumber\\%
\nonumber\\%
&&\sum_{\mu=x,y,z}|C_{\mu}^{(...\mathit{v}_{\mu}\!+\!1...)}|^2\exp[-\beta\hbar\Omega_{\mu}]|\mathit{v}_x,\mathit{v}_y,\mathit{v}_z\rangle\langle\mathit{v}_x,\mathit{v}_y,\mathit{v}_z|,%
\nonumber\\%
\label{a.4}%
\end{eqnarray}
where we additionally shifted up the vibrational sum on one unit in the second term, so the sum still starts from  $\mathit{v}_x+\mathit{v}_y+\mathit{v}_z\geq 1$. The obtained density operator describes a stationary mixed state which is, however, not expressed here by a Gibbs-type measure of a canonical ensemble.

In the total density matrix (\ref{a.2}) we can select the particular contribution of the dark-state
\begin{eqnarray}
\hat{\rho}^{(1)}&=&\exp\left\{\beta\left[{\cal F}(\beta)-\epsilon_{000}\right]\right\}%
\nonumber\\%
&\times&\left\{1+2\exp[-\beta\hbar\Omega_{\perp}]+\exp[-\beta\hbar\Omega_{\parallel}]\right\}%
\nonumber\\%
\nonumber\\%
&\times&|0,0,0\rangle\langle0,0,0|\times|\mathrm{s}\rangle\langle\mathrm{s}|%
\nonumber\\%
\nonumber\\%
&+&\ldots
\label{a.5}%
\end{eqnarray}
which is now enhanced by a factor coinciding with the oscillator's partition function cutoff up to the first excitation order with $\mathit{v}_x+\mathit{v}_y+\mathit{v}_z\leq 1$.

To justify this observation in general case let us calculate the enhancement factor after the second step of the protocol. To do this we have to keep only the following specific states in expansion (\ref{a.4}): $|1,0,0\rangle$ (repopulated from the states $|2,0,0\rangle$, $|1,1,0\rangle$ and $|1,0,1\rangle$), $|0,1,0\rangle$ (repopulated from $|1,1,0\rangle$, $|0,2,0\rangle$, and $|0,1,1\rangle$), and $|0,0,1\rangle$ (repopulated from $|1,0,1\rangle$, $|0,1,1\rangle$, and $|0,0,2\rangle$). The $C$-coefficients contributing to (\ref{a.4}) fulfil the following normalization conditions
\begin{eqnarray}
\lefteqn{\hspace{-2cm}|C_x^{(200)}|=1, \ \ \ \ \ |C_y^{(020)}|=1,\ \ \ \ \ |C_z^{(002)}|=1}%
\nonumber\\%
&&\hspace{-2cm}|C_x^{(110)}|^2+|C_y^{(110)}|^2=1%
\nonumber\\%
&&\hspace{-2cm}|C_x^{(101)}|^2+|C_z^{(101)}|^2=1%
\nonumber\\%
&&\hspace{-2cm}|C_y^{(011)}|^2+|C_z^{(011)}|^2=1.%
\label{a.6}%
\end{eqnarray}
The coefficients are not independent and cannot be taken as arbitrary parameters of the protocol since they are explicitly determined by the complete structure of the applied Raman pulse.

Let us clarify the expansion (\ref{a.5}) by showing those terms, which would be further transformed to the dark state at the second step of the protocol
\begin{eqnarray}
\hat{\rho}^{(1)}&=&\exp\left\{\beta\left[{\cal F}(\beta)-\epsilon_{000}\right]\right\}%
\nonumber\\%
&\times&\left\{1+2\exp[-\beta\hbar\Omega_{\perp}]+\exp[-\beta\hbar\Omega_{\parallel}]\right\}%
\nonumber\\%
\nonumber\\%
&&\times|0,0,0\rangle\langle0,0,0|\times|\mathrm{s}\rangle\langle\mathrm{s}|%
\nonumber\\%
\nonumber\\%
&+&\exp\left\{\beta\left[{\cal F}(\beta)-\epsilon_{100}\right]\right\}%
\nonumber\\%
&\times&\left\{|C_{x}^{(200)}|^2\exp[-\beta\hbar\Omega_{\perp}]+|C_{y}^{(110)}|^2\exp[-\beta\hbar\Omega_{\perp}]\right.
\nonumber\\%
&&\left.+|C_{z}^{(101)}|^2\exp[-\beta\hbar\Omega_{\parallel}]\right\}|1,0,0\rangle\langle 1,0,0|\times|\mathrm{s}\rangle\langle\mathrm{s}|%
\nonumber\\%
\nonumber\\%
&+&\exp\left\{\beta\left[{\cal F}(\beta)-\epsilon_{010}\right]\right\}%
\nonumber\\%
&\times&\left\{|C_{x}^{(110)}|^2\exp[-\beta\hbar\Omega_{\perp}]+|C_{y}^{(020)}|^2\exp[-\beta\hbar\Omega_{\perp}]\right.
\nonumber\\%
&&\left.+|C_{z}^{(011)}|^2\exp[-\beta\hbar\Omega_{\parallel}]\right\}|0,1,0\rangle\langle 0,1,0|\times|\mathrm{s}\rangle\langle\mathrm{s}|%
\nonumber\\%
\nonumber\\%
&+&\exp\left\{\beta\left[{\cal F}(\beta)-\epsilon_{001}\right]\right\}%
\nonumber\\%
&\times&\left\{|C_{x}^{(101)}|^2\exp[-\beta\hbar\Omega_{\perp}]+|C_{y}^{(011)}|^2\exp[-\beta\hbar\Omega_{\perp}]\right.
\nonumber\\%
&&\left.+|C_{z}^{(002)}|^2\exp[-\beta\hbar\Omega_{\parallel}]\right\}|0,0,1\rangle\langle 0,0,1|\times|\mathrm{s}\rangle\langle\mathrm{s}|%
\nonumber\\%
\nonumber\\%
&+&\ldots%
\label{a.7}%
\end{eqnarray}
After the round of both the Raman conversion and optical pumping cycles the selected terms would be transformed into the dark state and incorporated as one contribution. With taking into account (\ref{a.6}) we arrive at the following modification of the density matrix at the second step of the protocol
\begin{eqnarray}
\hat{\rho}^{(2)}&=&\exp\left\{\beta\left[{\cal F}(\beta)-\epsilon_{000}\right]\right\}%
\nonumber\\%
\nonumber\\%
&\times&\left\{1+2\exp[-\beta\hbar\Omega_{\perp}]+\exp[-\beta\hbar\Omega_{\parallel}]\right.%
\nonumber\\%
&&\left.+3\exp[-2\beta\hbar\Omega_{\perp}]+2\exp[-\beta\hbar\Omega_{\perp}-\beta\hbar\Omega_{\parallel}]\right.%
\nonumber\\%
&&\left.+\exp[-2\beta\hbar\Omega_{\parallel}]\right\}\times|0,0,0\rangle\langle0,0,0|\times|\mathrm{s}\rangle\langle\mathrm{s}|%
\nonumber\\%
\nonumber\\%
&+&\ldots%
\label{a.8}%
\end{eqnarray}
which can be equivalently written as
\begin{eqnarray}
\hat{\rho}^{(2)}&=&\exp\left\{\beta{\cal F}(\beta)\right\}%
\sum_{\begin{array}{c}\scriptstyle{\mathit{v_x,v_y,v_z}}\\ \scriptstyle{\mathit{v}_x+\mathit{v}_y+\mathit{v}_z\leq 2}\end{array}}\exp\left[-\beta\epsilon_{\mathit{v}_x\mathit{v}_y\mathit{v}_z}\right]%
\nonumber\\%
\nonumber\\%
&&\times|0,0,0\rangle\langle0,0,0|\times|\mathrm{s}\rangle\langle\mathrm{s}|%
\nonumber\\%
&+&\ldots%
\label{a.9}%
\end{eqnarray}
The remaining terms, indicated by ellipsis describe the population of the excited states of the oscillator, and the statistical distribution is not reproduced by the canonical Gibbs measure. The population of the excited states is suppressed, and the contribution of the dark state is enhanced by the partition function cutoff in the second order of the oscillator's excitation spectrum with $\mathit{v}_x+\mathit{v}_y+\mathit{v}_z\leq 2$.

Expanding the above arguments up to higher orders of the RSC-protocol we can justify the strategic result, reproduced by Eqs.~(\ref{2.6})-(\ref{2.9}) in the main text. It might seem that the above transformations could be revised for anharmonic potential as well. However we could point out that for preparing the Raman passage in a multilevel configuration one needs a reasonably limited number of the control pulses of different carrier frequencies, which can be provided for equidistant vibrational steps and would be difficult to do for highly excited anharmonic oscillator.

\section{The matrix elements contributed to (\ref{3.13}) and (\ref{3.14})}\label{Appendix_B}

\noindent As a representative example we consider here only those coupling coefficients in Eqs.~(\ref{3.13}) and (\ref{3.14}), which are responsible for the Raman passage with quenching of the vibration along $x$-direction. Other matrix elements of the effective Hamiltonian can be similarly constructed from the expressions, derived below, with simple change in the mode indices. For the same reason, we can associate $|b'\rangle$ and $|b\rangle$ with one type of states, which we further denote as $|b\rangle$. We specify all the matrix elements, contributing to the coupling coefficients, by the complete set of the system quantum numbers and then express them via the basic spectral parameters of the process.

For the coefficient, responsible for transferring the atom from state $|b\rangle$ to state $|m\rangle$ in Eq~(\ref{3.14}), we obtain
\begin{widetext}
\begin{eqnarray}
\lefteqn{\sum_{n}\frac{\Omega_{mn}^{(1)}\Omega_{nb}^{(0)}}{4\Delta_{n}}=\sum_{FM}\frac{1}{\hbar^2\Delta_{F}}}
\nonumber\\%
&&=\sum_{w_x,w_y,w_z}\langle F_{-}M_{-};\mathit{v}_x-1,\mathit{v}_y,\mathit{v}_z|(\mathbf{d}\cdot\mathbf{E}_1^{\ast})\mathrm{e}^{-i\mathbf{k}_1\cdot\mathbf{r}}|FM;w_x,w_y,w_z\rangle%
\langle FM;w_x,w_y,w_z|(\mathbf{d}\cdot\mathbf{E}_0)\mathrm{e}^{+i\mathbf{k}_0\cdot\mathbf{r}}|F_{+}M_{+};\mathit{v}_x,\mathit{v}_y,\mathit{v}_z\rangle%
\nonumber\\%
\nonumber\\%
&&=\sum_{FM}\frac{\langle F_{-}M_{-}|(\mathbf{d}\cdot\mathbf{E}_1^{\ast})|FM\rangle\langle FM|(\mathbf{d}\cdot\mathbf{E}_0)|F_{+}M_{+}\rangle}{\hbar^2\,\Delta_{F}}%
\times\langle\mathit{v}_x-1|\exp\left[i\frac{2}{\sqrt{3}}k_0x\right]|\mathit{v}_x\rangle,%
\label{b.1}%
\end{eqnarray}
\end{widetext}
where in the last line we made use of the completeness relation for the vibrational degrees of freedom for the atom excited to the upper state. In the definitions of the Rabi-frequencies the $\mathbf{E}_0$ and $\mathbf{E}_1$ are the vectors of complex amplitudes of the depopulating and the control modes respectively. The detuning $\Delta_n\equiv\Delta_F$ is specified by the total angular momentum of the upper state hyperfine sublevels. The exponent in the vibrational matrix element, apart selected to the last factor, contains only $x$-directed displacement of the atom's position from the trap origin, that shows that vibration quenching along $x$ direction is associated with switching on the $\omega_1$ control mode.

The selected matrix element can be evaluated as follows
\begin{eqnarray}
\lefteqn{\hspace{-1cm}\langle\mathit{v}_x-1|\exp\left[i\frac{2}{\sqrt{3}}k_0x\right]|\mathit{v}_x\rangle\approx \langle\mathit{v}_x-1|i\frac{2}{\sqrt{3}}k_0x|\mathit{v}_x\rangle}%
\nonumber\\%
&&=i\frac{2}{\sqrt{3}}\,k_0\,\sqrt{\frac{\hbar\mathit{v}_x}{2m_{\mathrm{A}}\Omega_{\perp}}}=i\frac{2}{\sqrt{3}}\,\eta_{\perp}\,\sqrt{\mathit{v}_x}%
\label{b.2}
\end{eqnarray}
where
\begin{equation}
\eta_{\perp} = k_0\,x_0 = k_0\,\sqrt{\frac{\hbar}{2m_{\mathrm{A}}\Omega_{\perp}}}
\label{b.3}
\end{equation}
is the so called Lamb-Dicke parameter (factor) and $x_0=\sqrt{\hbar/2m_{\mathrm{A}}\Omega_{\perp}}$ is the spread of the zero-point oscillator wave-function, and $m_{\mathrm{A}}$ is the atomic mass. In the above estimate we have assumed that $\eta_{\perp}\sqrt{\mathit{v}_x}\ll 1$. Otherwise the precise evaluation of the above matrix element would be needed, see \cite{Winelend98}. Nevertheless just validity of such a strong inequality is a key requirement for applicability of the RSC-protocol itself. Remind that at the optical pumping stage of the protocol it is crucially important that the atom would preserve its vibrational mode in interaction with the pump light, which is provided by a small value of the Lamb-Dicke factor.

The coupling coefficient responsible for the inverse process contributed to Eq.~(\ref{3.13}), which repopulates the atom back to the Zeeman states of the upper hyperfine sublevel, can be expressed in a similar way
\begin{widetext}
\begin{eqnarray}
\lefteqn{\sum_{n}\frac{\Omega_{bn}^{(0)}\Omega_{nm}^{(1)}}{4\Delta_{n}}=\sum_{FM}\frac{1}{\hbar^2\Delta_{F}}}
\nonumber\\%
&&=\sum_{w_x,w_y,w_z}\langle F_{+}M_{+};\mathit{v}_x,\mathit{v}_y,\mathit{v}_z|(\mathbf{d}\cdot\mathbf{E}_0^{\ast})\mathrm{e}^{-i\mathbf{k}_0\cdot\mathbf{r}}|FM;w_x,w_y,w_z\rangle%
\langle FM;w_x,w_y,w_z|(\mathbf{d}\cdot\mathbf{E}_1)\mathrm{e}^{+i\mathbf{k}_1\cdot\mathbf{r}}|F_{-}M_{-};\mathit{v}_x-1,\mathit{v}_y,\mathit{v}_z\rangle%
\nonumber\\%
\nonumber\\%
&&=\sum_{FM}\frac{\langle F_{+}M_{+}|(\mathbf{d}\cdot\mathbf{E}_0^{\ast})|FM\rangle\langle FM|(\mathbf{d}\cdot\mathbf{E}_1)|F_{-}M_{-}\rangle}{\hbar^2\,\Delta_{F}}%
\times\langle\mathit{v}_x|\exp\left[-i\frac{2}{\sqrt{3}}k_0x\right]|\mathit{v}_x-1\rangle%
\label{b.4}%
\end{eqnarray}
\end{widetext}
where
\begin{eqnarray}
\lefteqn{\hspace{-1cm}\langle\mathit{v}_x|\exp\left[-i\frac{2}{\sqrt{3}}k_0x\right]|\mathit{v}_x-1\rangle\approx -\langle\mathit{v}_x|i\frac{2}{\sqrt{3}}k_0x|\mathit{v}_x-1\rangle}%
\nonumber\\%
&&=-i\frac{2}{\sqrt{3}}\,k_0\,\sqrt{\frac{\hbar\mathit{v}_x}{2m_{\mathrm{A}}\Omega_{\perp}}}=-i\frac{2}{\sqrt{3}}\,\eta_{\perp}\,\sqrt{\mathit{v}_x}%
\label{b.5}
\end{eqnarray}
Other terms associated with the quenching of the vibrations along $y$ and $z$ directions can be straightforwardly written with a simple modification of the above equations by substituting mode index $1\to 2,\,3$, vibrational quantum number $\mathit{v}_x\to \mathit{v}_y,\,\mathit{v}_z$, and oscillator frequency $\Omega_{\perp}\to\Omega_{\perp},\;\Omega_{\parallel}$ and respectively the Lamb-Dicke factor $\eta_{\perp}\to\eta_{\perp},\;\eta_{\parallel}$.

Other terms in the right-hand side of the system (\ref{3.13}) and (\ref{3.14}) contain the coefficients, which are diagonal in the basis of the oscillator states and are independent on the vibrational quantum numbers. The interaction solely with the depopulating mode is expressed by the following coefficients
\begin{widetext}
\begin{eqnarray}
\lefteqn{\sum_{n}\frac{|\Omega_{nb}^{(0)}|^2}{4\Delta_{n}}=\sum_{FM}\frac{1}{\hbar^2\Delta_{F}}}
\nonumber\\%
&&=\sum_{w_x,w_y,w_z}\langle F_{+}M_{+};\mathit{v}_x,\mathit{v}_y,\mathit{v}_z|(\mathbf{d}\cdot\mathbf{E}_0^{\ast})\mathrm{e}^{-i\mathbf{k}_0\cdot\mathbf{r}}|FM;w_x,w_y,w_z\rangle%
\langle FM;w_x,w_y,w_z|(\mathbf{d}\cdot\mathbf{E}_0)\mathrm{e}^{+i\mathbf{k}_0\cdot\mathbf{r}}|F_{+}M_{+};\mathit{v}_x,\mathit{v}_y,\mathit{v}_z\rangle%
\nonumber\\%
\nonumber\\%
&&=\sum_{FM}\frac{\langle F_{+}M_{+}|(\mathbf{d}\cdot\mathbf{E}_0^{\ast})|FM\rangle\langle FM|(\mathbf{d}\cdot\mathbf{E}_0)|F_{+}M_{+}\rangle}{\hbar^2\,\Delta_{F}}%
\label{b.6}%
\end{eqnarray}
and
\begin{eqnarray}
\lefteqn{\sum_{n}\frac{|\Omega_{nm}^{(0)}|^2}{4(\Delta_{n}-\Delta_{\mathrm{hpf}})}=\sum_{FM}\frac{1}{\hbar^2(\Delta_{F}-\Delta_{\mathrm{hpf}})}}
\nonumber\\%
&&\hspace{-0.5cm}=\!\!\sum_{w_x,w_y,w_z}\!\!\langle F_{-}M_{-};\ldots,\mathit{v}_{\mu}\!-\!1,\ldots|(\mathbf{d}\!\cdot\!\mathbf{E}_0^{\ast})\mathrm{e}^{-i\mathbf{k}_0\cdot\mathbf{r}}|FM;w_x,w_y,w_z\rangle%
\langle FM;w_x,w_y,w_z|(\mathbf{d}\!\cdot\!\mathbf{E}_0)\mathrm{e}^{+i\mathbf{k}_0\cdot\mathbf{r}}|F_{-}M_{-};\ldots,\mathit{v}_{\mu}\!-\!1,\ldots\rangle%
\nonumber\\%
\nonumber\\%
&&\hspace{-0.5cm}=\sum_{FM}\frac{\langle F_{-}M_{-}|(\mathbf{d}\cdot\mathbf{E}_0^{\ast})|FM\rangle\langle FM|(\mathbf{d}\cdot\mathbf{E}_0)|F_{-}M_{-}\rangle}{\hbar^2(\Delta_{F}-\Delta_{\mathrm{hpf}})}%
\label{b.7}%
\end{eqnarray}
\end{widetext}
which are both insensitive to the vibrational motion.

Each amplitude $\mathbf{E}_0$ and $\mathbf{E}_j$ can be factorized as $\mathbf{E}_0=\mathbf{e}^{(0)}{\cal E}_0$ and $\mathbf{E}_j=\mathbf{e}^{(j)}{\cal E}_j$ (no sum) and give us a set of the unit polarization vectors separated from the scalar part of the complex field amplitudes. In order to find the above coefficients in the closed form we have to evaluate the matrix element for the product $\textbf{d}\cdot\textbf{e}$ where $\mathbf{e}$ can be any of the mode polarization vectors. The tricky point is that in the above equations all the vector components are defined in respect to the reference frame associated with the depopulating beam. So the projections of $\mathbf{e}^{(0)}$ and $\mathbf{e}^{(j)}$ have to be precisely specified and connected with the considered experimental geometry shown in the diagrams of Figs.~\ref{fig2} and \ref{fig4}.

The atomic dipole moment is a real vector and physical observable, but it is convenient to express this quantity via its complex spherical components. The complex basis set of spherical unit vectors is given by
\begin{eqnarray}
\mathbf{e}_{0}&=&\mathbf{e}_{z}%
\nonumber\\
\mathbf{e}_{\pm 1}&=&\mp (\mathbf{e}_{x}\pm i\mathbf{e}_{y})/\sqrt{2}%
\label{b.8}
\end{eqnarray}
Then the spherical components of the dipole operator are given by its projections on these vectors
\begin{eqnarray}
d_q&=&\textbf{d}\cdot\textbf{e}_q%
\nonumber\\
d_{0}&=&d_{z}%
\nonumber\\
d_{\pm 1}&=&\mp (d_{x}\pm id_{y})/\sqrt{2}%
\label{b.9}
\end{eqnarray}
and their angular dependence is equivalent to $Y_{1q}(\theta,\phi)$ spherical functions.

The basic matrix element of the dipole operators is off-diagonal in the basis of the ground and excited atomic states specified by the quantum numbers of total angular momentum and its projection
\begin{equation}
\left(\textbf{d}\cdot\textbf{e}_q\right)_{nm}\ \equiv\ \langle F,M|d_q|F_0,M_0\rangle%
\label{b.10}
\end{equation}
The transition matrix element of an atomic dipole operator can be evaluated with the aid of the Wigner-Eckart theorem and can be factorized in the following product
\begin{equation}
\langle F,M|d_q|F_0,M_0\rangle=\frac{\langle F\!\parallel\! d\!\parallel\! F_0\rangle}{\sqrt{2F+1}}\;C^{FM}_{F_0M_0\,1q}%
\label{b.11}
\end{equation}
where $C_{\ldots\,\ldots}^{\ldots}$ is the Clebsch-Gordan coefficient and factor $\langle F\!\parallel\! d\!\parallel\! F_0\rangle$ is the reduced matrix element of the dipole (vector) operator , see \cite{LaLfIII}.

The quantum numbers $F,M$ and $F_0,M_0$ are the values of the total angular momenta for the composition of electronic (orbital and spin) and nuclear (spin) states into a coupled hyperfine state.  In the decoupled basis the dipole operator does not affect the nuclear subsystem. In this case it is convenient to eliminate the nuclear subsystem according to the weakness of the hyperfine interaction with respect to the spin-orbital interaction. Omitting the derivation details we reproduce here the final result. The reduced matrix element of the dipole operator can be factorized as follows
\begin{eqnarray}
\langle F\!\parallel\! d\!\parallel\! F_0\rangle &=& (-)^{F_0+J+I-1}\left[(2F+1)(2F_0+1)\right]^{1/2}%
\nonumber\\%
&&\times\left\{\begin{array}{ccc} S & I & F_0\\ F & 1 & J \end{array}\right\}%
\langle J\!\parallel\! d\!\parallel\! S\rangle %
\label{b.12}
\end{eqnarray}
where the factor $\langle J\!\parallel\! d\!\parallel\! S\rangle$ performs the reduced matrix element when the nuclear subsystem is completely ignored. Here $J$ is the total (spin and orbital) angular momentum of the excited state and $S\equiv J_0=1/2$ is the electronic spin coinciding with the total angular momentum of the ground state. The table-factor in curly braces is so called  $6j$-symbol appearing due to decomposition of the coupled state in the decoupled basis of the electronic and nuclear spin subsystems, see \cite{LaLfIII}.

The performed factorization of the transition matrix element for an atomic dipole operator allows us to express it by an experimentally measurable parameter, namely, by the spontaneous radiation decay rate, which is given by
\begin{equation}
\gamma_J\ =\ \frac{4\omega_{J0}^3}{3\hbar c^3}\;\frac{|\langle J\!\parallel\! d\!\parallel\! S\rangle|^2}{2J+1}\sim\gamma%
\label{b.13}
\end{equation}
where $\omega_{J0}$ is the transition frequency for either $J=1/2$ ($D_1$-line) or $J=3/2$ ($D_2$-line). In reality the decay rate $\gamma_J$ is weakly sensitive to the fine structure splitting in the upper state such that it is practically the same for both the lines. Thus the expressions (\ref{b.11})-(\ref{b.13}) allow us to scale all the set of the transition matrix elements for an atomic dipole via one and well known experimental parameter $\sqrt{\gamma}$. But in the case of equations (\ref{3.13}) and (\ref{3.14}) it seems more natural to incorporate the reduced dipole moment and the field amplitudes into the set of reduced Rabi-frequencies given by
\begin{eqnarray}
\Omega^{(0)}&=& 2|\langle J\!\parallel\! d\!\parallel\! S\rangle|{\cal E}_0%
\nonumber\\%
\Omega^{(j)}&=& 2|\langle J\!\parallel\! d\!\parallel\! S\rangle|{\cal E}_j%
\label{b.14}%
\end{eqnarray}
which can be scaled by the decay rate $\gamma$.

\bibliographystyle{apsrev4-1}
\bibliography{references}

\end{document}